\documentclass[a4paper,fleqn,usenatbib,useAMS]{mnras}

\usepackage{newtxtext,newtxmath}
\usepackage{graphicx}

\usepackage[T1]{fontenc}
\usepackage{ae,aecompl}

\usepackage{amsmath}	
\usepackage[export]{adjustbox}

\newcommand{\pcm}{\,cm$^{-2}$ }	
\newcommand{\cmmo}{\,cm$^{-1}$ }	
\newcommand{\pcmc}{\,cm$^{-3}$ } 
\newcommand{\kmps}{\,km s$^{-1}$ } 

\title[An accurate set of H$_3$O$^+ -$ H$_2$ collisional rate coefficients]{An accurate set of H$_3$O$^+ -$ H$_2$ collisional rate coefficients for non-LTE modelling of warm interstellar clouds}

\author[S. Demes et al.]{
\thanks{Corresponding author's e-mail: sandor.demes@univ-rennes1.fr} S\'{a}ndor Demes$^{1}$, \thanks{E-mail: francois.lique@univ-rennes1.fr} Fran\c{c}ois Lique$^{1}$, Alexandre Faure$^{2}$, Floris F. S. van der Tak$^{3}$
\\
$^{1}$ Univ Rennes, CNRS, IPR (Institut de Physique de Rennes) - UMR 6251, F-35000 Rennes, France\\ $^{2}$ IPAG, Universit\'{e} Grenoble Alpes \& CNRS, CS 40700, F-38058 Grenoble, France\\ $^{3}$ SRON Netherlands Institute for Space Research \& Kapteyn Astronomical Institute, University of Groningen, 9747 AD Groningen, The Netherlands\\
}

\date{Accepted xxxx/yy/zz}

\pubyear{2022} 

\begin{document}
\label{firstpage}
\pagerange{\pageref{firstpage}--\pageref{lastpage}}
\maketitle

\begin{abstract}
Hydronium (H$_3$O$^+$) was first detected in 1986 in interstellar molecular clouds. It was reported in many galactic diffuse and dense regions, as well as in extragalactic sources. H$_3$O$^+$ plays a major role both in interstellar oxygen and water chemistry. However, despite the large number of H$_3$O$^+$ observations, its collisional excitation was investigated only partially. In the present work we study the state-to-state rotational de-excitation of {\it ortho-} and {\it para-}H$_3$O$^+$ in collisions both with {\it ortho-} and {\it para-}H$_2$. The cross sections are calculated within the close-coupling formalism using a highly accurate potential energy surface developed for this system. The rate coefficients are computed up to $300$ K kinetic temperature. Transitions between the lowest 21 rotation-inversion states were studied for {\it para-}H$_3$O$^+$, and the lowest 11 states for {\it ortho-}H$_3$O$^+$, i.e. all levels with rotational energies below 430 K ($\sim 300$ cm$^{-1}$) are considered (up to $j\leq5$). In order to estimate the impact of the new rate coefficients on the astrophysical models for H$_3$O$^+$, radiative transfer calculations were also carried out. We have examined how the new collisional data affect the line intensities with respect to older data previously used for the interpretation of observations. By analysing all detected transitions we find that our new, accurate rate coefficients have a significant impact (typically within a factor of 2) on radiation temperatures, allowing more accurate estimation of column densities and relative abundances of hydronium, especially in warm molecular clouds, paving the path towards better interpretation of interstellar water and oxygen chemistry.

\end{abstract}




\section{Introduction}\label{sec:Intro}

The presence of the hydronium cation (H$_3$O$^+$) in various interstellar molecular clouds was reported by many observational research papers, including its prediction in a fundamental work by \citet{Herbst1973} followed by the first nearly simultaneous observations in 1986 by \citet{wootten86} and \citet{hollis86}, then the confirming detections in the 90s of the past century \citep{wootten91,Phillips1992,timmermann96}, and later in early 2000s by \citet{van_der_tak_06,VanderTak2008}. In the 2010s there were several new observations of H$_3$O$^+$ reported in different interstellar environments \citep{benz10,Gerin2010,Gonzalez2013,lis14,Indriolo2015, benz16,ando17}, while the most recent studies were published by \citet{martin21} and \citet{holdship22}. The existing observational detections of hydronium will be reviewed thoroughly in Section~\ref{sec:ObsLit}. Chemical models involving hydronium were also developed intensively, as one can see, for example, in the works of \citet{Sternberg1995,Goicoechea2001,Hollenbach2012,VanDishoeck2013,faure19}, which all reported that it plays a key role in oxygen and water chemistry of the interstellar medium (ISM) nominating it as a crucial chemical species in interstellar ion-molecule reaction schemes. The relative abundances of H$_3$O$^+$ and H$_2$O may be also used to measure the ionization rates of molecular clouds both in the Milky Way \citep{van_der_tak_06} as well as in external galaxies \citep{VanderTak2008, VanDerTak2016}. According to \citet{Phillips1992}, measuring the abundance of the H$_3$O$^+$ cation can indirectly provide the abundance of interstellar water. To quantitatively determine its role in different astrophysical environments, however, the details of both its radiative and collisional excitation should be taken into account in addition to its formation and destruction paths.

The formation mechanisms of H$_3$O$^+$ in molecular clouds as well as its dissociative recombination pathways were briefly summarised in our previous paper (\citet{Demes2021}, referred as Paper~II hereinafter). The relevant details about the chemistry of hydronium are discussed by \citet{VanDishoeck2013} and \citet{Hollenbach2012}, while its dissociative recombination is carefully studied by \citet{Jensen2000}. The high-resolution spectroscopy measurements for H$_3$O$^+$ are collected and analyzed by \citet{SYu2009}. While the spectroscopy of the pure rotational levels of hydronium is rather complete, its hyperfine structure remains unresolved \citep{SYu2009}.
For molecules with non-zero nuclear spin, hyperfine splitting is usually only observable if they contain a nucleus with spin $I \ge 1$, such as $^{17}$O or D \citep{vanderTak09}. In interstellar clouds, such splitting may be observed for the H$_3$O$^+$ isotopologues H$_3^{17}$O$^+$ and H$_2$DO$^+$. Its electron-impact excitation was studied also earlier \citep{Faure2003}. The collisional excitation studies of hydronium are very limited however. The first, approximated collisional cross section were provided by \citet{Offer1992}, which has given a basis then for interpreting the observations of H$_3$O$^+$ in both dense and diffuse interstellar clouds. However the authors did not report rate coefficients, so their data were scaled in combination with NH$_3$ collisional data of \citet{danby88}. Later \citet{ElHanini2019} studied the rotational excitation of hydronium by helium atoms (as a template for H$_2$), but for a limited energy and temperature range (up to $50$ K). Recently we have computed state-to-state collisional rate coefficients up to 100 K for the rotational excitation of both $ortho$- and $para$-H$_3$O$^+$ in collision with $para$-H$_2$ \citep{Demes2021}. A detailed discussion about the previous collisional excitation studies \citep{Offer1992, ElHanini2019} was already given in this work, including a comparative analysis of the cross sections and rate coefficients.

It is worth to emphasise again that, according to the authors of several observational studies (see, for example, the works by \citet{wootten91,VanderTak2008}) reliable and precise collisional rate coefficients are needed to correctly interpret the astrophysical observations of H$_3$O$^+$ in ISM clouds. Without such data the molecular abundances could be only approximated, assuming local thermodynamic equilibrium (LTE), which is usually not a good approximation for typical interstellar conditions \citep{Roueff2013}.

In the present work we provide a complete set of rotational excitation cross sections and thermal rate coefficients for the H$_3$O$^+ -$ H$_2$ collisional system, which are calculated based on an accurate 5-dimensional interaction potential (\citet{Demes2020}, referred as Paper~I hereinafter). The new collisional data presented here have several crucial improvements compared to Paper~II: the calculations are extended to {\it ortho-}H$_2$, and we compute accurate rate coefficients up to 300 K (over 100 K in Paper~II). To demonstrate the impact of the new collisional data on radiative transfer calculations are provided as well and their results obtained with different sets of rate coefficients are compared.

The paper is organised as follows: In Section~\ref{sec:ObsLit} the literature of the H$_3$O$^+$ observations is reviewed in a chronological order. In Section~\ref{sec:ModMeth} the details of the interaction potential as well as the scattering calculations are presented. The state-to-state collisional data for the rotational de-excitation of H$_3$O$^+$ by H$_2$ are discussed in Section~\ref{sec:Results}, where subsection~\ref{sec:CSs} introduces the cross section and subsection~\ref{sec:RCs} presents the thermal rate coefficients, respectively. In Section~\ref{sec:Excit} we also report radiative transfer calculation results, while our concluding remarks are drawn in Section~\ref{sec:Conclusions}.

\section{Observational detections of H$_3$O$^+$}\label{sec:ObsLit}

The first prediction of the presence of H$_3$O$^+$ in dense molecular clouds of the Orion/KL and Sgr B2 regions (where the approximate hydrogen density is $\sim 10^6$ \pcmc) was given by \citet{Herbst1973}. More than a decade later \citet{wootten86} and \citet{hollis86}  almost simultaneously reported the first detections. Both groups used the NRAO 12 m telescope and both measured the 307.2 GHz emission line in the OMC-1 and Sgr B2 sources. The confirming detection was reported soon by \citet{wootten91} in the same environments with the identification of the $3_2^+ \rightarrow 2_2^-$ transition of {\it ortho}-H$_3$O$^+$ at 364 GHz (note that the rotational states of H$_3$O$^+$ are denoted by $j_k^{\epsilon}$, where $j$ is the total angular momentum, $k$ is its projection on the $C_3$ main rotational axis, and $\epsilon=\pm$ is the inversion symmetry index, see \cite{Rist1993} for more details). In these first observations the authors reported kinetic temperatures above 70 K with an H$_2$ density of about $10^6$ \pcmc and they also emphasise the importance of a non-local thermodynamic equilibrium (non-LTE) analysis, since the low excitation temperature of H$_3$O$^+$ are poorly described in an LTE approach (which was the only available model for H$_3$O$^+$ at that time due to the lack of collisional data). Later \citet{Phillips1992} reported a more clear detection of three - 396, 364 and 307 GHz - submillimeter lines in more than 10 galactic sources, besides the spectroscopically confusing OMC-1 and Sgr B2 clouds, the G34.3+0.15 and W51 clouds as well as different high-mass star formation regions, for example the W3 IRS 5 cloud. Most of them are exhibiting high-density ($n_\mathrm{H_2} \approx 10^6-10^7$ \pcmc) and high temperature ($T_\mathrm{kin} \gtrsim 50$ K) conditions. The authors found that the typical H$_3$O$^+$ abundance is $10^{-10}-10^{-9}$ in these regions. The $4_3^- \rightarrow 3_3^+$ far-infrared emission line of H$_3$O$^+$ at $69.524$ $\mu$m (4312 GHz) was measured soon after by \citet{timmermann96} in the Orion BN-IRc2 region with NASA's Kuiper Airborne Observatory. The authors reported it as a possible  H$_3$O$^+$ detection, which might originate from very dense clumps. They also reported high hydrogen densities ($\gtrsim 5 \times 10^8$ \pcmc) and temperatures ($\gtrsim 100$ K) in this environment. For their model and analysis the authors used scaled, approximated collisional rate coefficients based on the work of \citet{Offer1992}, and they estimated the column densities of H$_3$O$^+$ in the range from $9 \times 10^{13}$ \pcm (under LTE conditions and 300 K) up to $7 \times 10^{16}$ \pcm (under non-LTE conditions and $T = 100$ K).

Ten years later \citet{van_der_tak_06} detected strong emission in the H$_3$O$^+$ 364 GHz and 307 GHz lines both in the M core and OH envelope of the Sgr B2 region with the APEX telescope. The authors reported column densities in the range between $3.7 \times 10^{15}$ and $1.36 \times 10^{16}$ \pcm in the envelope and in the core, respectively and they found the hydronium abundance to be $\sim 3 \times 10^{-9}$ relative to H$_2$ and the H$_3$O$^+$/H$_2$O ratio to be about $1/50$ in this region. The physical conditions in the M core are $T = 200$ K and $n_\mathrm{H_2} = 10^7$ \pcmc, while in the low-density OH envelope the temperature is about $60$ K only and the H$_2$ density is $10^6$ \pcmc. Later hydronium was detected in extragalactic environments as well by \citet{VanderTak2008}. Observations with the JCMT targeted the well-known 364 GHz emission line of the $3_2^+ \rightarrow 2_2^-$ transition in the centers of M 82 and Arp 220, where the expected kinetic temperature is $\sim 100$ K and the hydrogen density is at least $10^5$ \pcmc. With an estimated $para$-H$_3$O$^+$ column density of $0.7-1.3 \times 10^{13}$ \pcm for Arp 220 and $3.0-5.5 \times 10^{13}$ \pcm for M 82, the authors found that the typical hydronium abundance is $2-10 \times 10^{-9}$ relative to H$_2$. For radiative transfer modelling of these environments both in the work of \citet{van_der_tak_06} and \citet{VanderTak2008} the scaled collisional data were derived from \citet{Offer1992} and \citet{danby88}.

The W3 IRS 5 cloud was later studied by \citet{benz10}, searching for the major hydrides in this region of the ISM. The authors detected several rotational transitions of H$_3$O$^+$ in this high-mass star-forming region using HIFI on the Herschel Space Observatory. Emissions from the $4_3^+ \rightarrow 3_3^-$ (1031.3 GHz), $4_2^- \rightarrow 3_2^+$ (1069.8 GHz), $6_2^+ \rightarrow 6_2^-$ (1454.6 GHz) and $2_1^+ \rightarrow 2_1^-$ (1632.1 GHz) $ortho$- and $para$-H$_3$O$^+$ transitions were measured. The authors found the column density is typically about $10^{14}$ \pcm, and the corresponding hydronium abundance is $4.2(\pm1) \times 10^{-10}$ relative to atomic hydrogen. The gas density of the environment exceeds  $10^7$ \pcmc,  and the rotational temperature is about $205 - 285$ K (the gas temperature can be even higher). In parallel a weak 984 GHz absorption line of H$_3$O$^+$ was also detected with the HIFI/{\it Herschel} along the massive star-forming region G10.6-0.4 (W31C) \citep{Gerin2010}. The authors derived the hydronium column density $N$(H$_3$O$^+)$ $\sim 4 \times 10^{13}$ \pcm  for the $V_\mathrm{LSR}$ velocity range of $7-45$ \kmps . According to \citet{Gerin2010} in this diffuse cloud the hydrogen density is only about $10^2-10^3$ \pcmc and the temperature can be as low as 5 K.

\citet{Gonzalez2013} also reported excited hydronium in the extragalactic NGC 4418 and Arp 220 regions based on {\it Herschel}/PACS observations.  The column density of H$_3$O$^+$ is found to be about $(0.9-2.7) \times 10^{16}$ \pcm in Arp 220, and the observed pure inversion, metastable lines have a very high rotational temperature of $\sim500$ K. The authors characterise the source as a relatively low density ($\gtrsim 10^4$ \pcmc) interclump medium. They presented the detection of a series of lines of hydronium in the $70-180$ $\mu$m spectral range, involving several states with $j=2-5$ rotational quantum numbers and above. It is also shown by \citet{Gonzalez2013}  that the states with $K > 1$ projection quantum numbers can be populated only through collisions or via molecular formation in high-lying levels. The authors still used scaled rate coefficients based on the data from \citet{Offer1992} and \citet{danby88} for non-LTE modelling of the H$_3$O$^+$ excitation. In the NGC 4418 source the work reports $(5-8) \times 10^{15}$ \pcm column densities and $T_\mathrm{gas} \sim150$ K for the transitions from lower $K \leq 5$ as well as $T_\mathrm{gas} \sim500$ K for the ones from higher rotational states . Just a year later \citet{lis14} presented the results of {\it Herschel} observations of the $6_6^- \rightarrow 6_6^+$ and $9_9^- \rightarrow 9_9^+$ inversion transitions of hot hydronium toward the diffuse Sagittarius B2(N) and W31C sources, with estimated column densities of $7 \times 10^{14}$ \pcm and $1.2 \times 10^{13}$ \pcm, respectively. The authors report rotational temperatures of about $500$ K and $380$ K, so the kinetic temperature can reach $400-600$ K. The next {\it Herschel} observation of H$_3$O$^+$ was presented by \citet{Indriolo2015} for diffuse interstellar clouds along 20 Galactic sight lines toward bright submillimeter continuum sources. The authors measured the $1_1^- \rightarrow 1_1^+$ (1655.8 GHz) transition in absorption and the $0_0^- \rightarrow 1_0^+$ (984.7 GHz) line in emission and they derived column densities in the range from $3 \times 10^{14}$ up to $1.66 \times 10^{13}$ \pcm depending on the source. The authors suggested a kinetic temperature of $100$ K, which represents the average temperature in diffuse clouds, and they considered hydrogen densities as low as 35 \pcmc. Another observation by {\it Herschel} \citep{benz16} reported the $4_3^+ \rightarrow 3_3^-$ (1031.3 GHz) emission line of $ortho$-H$_3$O$^+$ in several young stellar objects. The authors considered a wide range of gas densities ($10^4-10^6$ \pcmc) and kinetic temperatures (from 10 K up to several hundreds of K).

Emission from the $3_2^+ \rightarrow 2_2^-$ line of H$_3$O$^+$ (364 GHz) was detected again recently by \citet{ando17} in the starburst galaxy NGC 253, based on ALMA Band 7 observations. According to the authors, this is a dense ($n_\mathrm{H_2} \geq 10^6$ \pcmc) and hot environment with typical kinetic temperatures $\geq 100$ K. The latest observations of hydronium were reported by \citet{holdship22} within the ALMA Comprehensive High-Resolution Extragalactic Molecular Inventory. The research has targeted the central molecular zone of the extragalactic NGC 253 source, studying the impact of cosmic-ray ionization rates on the  abundance ratio of  H$_3$O$^+$ and SO. The authors estimated the fractional abundance of hydronium to be about $5 \times 10^{-10} - 10^{-9}$. They varied the gas density and temperature in a wide range, respectively from $10^4$ to $10^7$ \pcmc and from $50$ up to $300$ K. Their analysis was performed based on the well-studied 364 GHz ($3_2^+ \rightarrow 2_2^-$) and 307 GHz ($1_1^+ \rightarrow 2_1^+$) transitions of H$_3$O$^+$, leading to a conclusion that the H$_2$ density most likely vary between $0.4-9.8 \times 10^5$ \pcmc depending on the source.

\section{Methods}\label{sec:ModMeth}
In Paper~II we described the relevant approach of the scattering calculations. Here we briefly summarise the most important points and provide the details for the new calculations that are presented in this paper.

\subsection{Potential energy surface}\label{sec:PES}
The scattering calculations were performed using our recent 5D rigid-rotor potential energy surface (Paper~I) for the H$_3$O$^+ -$ H$_2$ interaction. This potential was calculated by the rigorously tested and very accurate CCSD(T)-F12 (explicitly correlated coupled-cluster theory with singles and doubles with perturbative corrections for triple excitations) {\it ab initio} theory with a moderate-size aug-cc-pVTZ (augmented correlation-consistent polarised valence-triple-$\zeta$) basis set. This combination of method + basis set ensures the necessary high quality of the PES with a reasonable computational cost.

The collisional system is defined in a Jacobi coordinate system using the molecular frame (body-fixed) representation. Consequently, the center of the coordinate system is set in the center of mass (c.o.m.) of the target H$_3$O$^+$ cation. One spatial parameter ($R$) defines the distance between the colliders, while four spherical angles describe the position and relative orientation of the H$_2$ projectile with respect to the origin of the coordinate system. The H$_2$ bond length was set at $r_\mathrm{H-H} = 1.44874$ a$_0$ \citep{Bubin2003}, and for H$_3$O$^+$ we used the experimental bond properties by \citet{Tang1999}: $r_\mathrm{O-H} = 1.8406$ a$_0$ and $\alpha_\mathrm{H-O-H} = 113.6^{\circ}$.

The {\it ab initio} PES was fitted using a standard least square procedure, resulting in a set of 208 radial expansion functions for each $R$, with anisotropies up to $l_1=16$ and $l_2=4$. The root mean square (rms) residual was found to be lower than 1~\cmmo in the long-range of the potential and also in the well region (which is about 1887.2 \cmmo deep). The rms error on the expansion coefficients was also found to be smaller than 1 \cmmo in these regions of the PES. A cubic spline interpolation of the analytical coefficients was performed using for distances between $R=4-30$ a$_0$ and it was smoothly connected then to standard extrapolations using the switch function proposed by \citet{Valiron2008}. Since there is no experimental observables for the H$_3$O$^+ -$ H$_2$ system, we cannot assess the accuracy of our interaction potential. However, from similar PES calculations by \citet{faure16,pirlot21,godard22}, where the accuracy of the potential was estimated based on a comparison with experimental dissociation and/or bound states energies, we expect that its overall accuracy is at the wavenumber level of accuracy and is always better than 10 per cents. For full details about the interaction potential and the analytical fit, see Paper~I.

We used the following units throughout this paper (until otherwise noted): atomic units (a$_0$) for distances (1~a$_0$ = 1~bohr $\approx 5.29177 \times 10^{-9}$ cm), wavenumbers (\cmmo) for energies (1~\cmmo $\approx 1.4388$ K).

\subsection{Scattering calculations}\label{sec:ScatCalc}

We have calculated the state-to-state rotational de-excitation cross sections and thermal rate coefficients for the collision of {\it ortho-} and {\it para-}H$_3$O$^+$ with {\it para-}H$_2$ in our previous work \citep{Demes2021}. This work is an extension of the previous one with several crucial improvements. First, we have studied all the possible nuclear spin isomers, so the calculations are extended to collisions with {\it ortho-}H$_2$ as well, which is the dominant collider in interstellar clouds with temperatures above 100 K. Second, we have significantly increased the interval of collision energies for the cross section calculations, which allowed us to compute accurate rate coefficients up to 300 K. These improved collisional data allows one to provide a precise non-LTE modelling for warm molecular clouds involving H$_3$O$^+$ (see, for example, the observations by \citet{VanderTak2008} at several hundred Kelvins).

In current work we determined the cross sections by the full quantum (close-coupling, CC) approach using the \texttt{HIBRIDON} scattering code \citep{Manolopoulos86,Millard87}. They were computed up to 1700 \cmmo total energies for H$_3$O$^+$ collisions with {\it o-}H$_2$ and up to 1500 \cmmo total energies for collisions with {\it p-}H$_2$. The {\it o-}H$_3$O$^+$ spin species is characterised with $k=3n$ quantum numbers (where $n$ are integer numbers), while all other $k$ quantum numbers refer to {\it p-}H$_3$O$^+$.

We computed all inelastic cross sections for the transitions between rotational levels with an internal energy of $\leq430$ K ($ \sim 300$ cm$^{-1}$), including states up to $5_2^+$ for {\it p-}H$_3$O$^+$ and $5_3^+$ for {\it p-}H$_3$O$^+$. For the full list of the rotational states involved for H$_3$O$^+$, see Table~\ref{tab:rot-levels} in Appendix~\ref{app:apdixA}. Both the rotational basis ($j_\mathrm{max}$) and the maximum total angular momentum ($J_\mathrm{tot}$) were selected depending on the nuclear spin symmetry and collision energy. Their particular values are set based on preliminary convergence test calculations with the following convergence-threshold criteria: maximum 1\% mean deviation for $j_\mathrm{max}$ and 0.01\% mean deviation for  $J_\mathrm{tot}$. As discussed in Paper~II, in order to adequately describe the strong resonances in the cross sections, a very small step size ($E_\mathrm{step} = 0.1$ \cmmo) was chosen at low collision energies, which was gradually increased then. The values of the $j_\mathrm{max}$, $J_\mathrm{tot}$ and $E_\mathrm{step}$ parameters, which were used in the calculations for the particular total energy intervals from $E_\mathrm{init}$ to $E_\mathrm{fin}$ are listed in Tables~\ref{tab:conv-param-pH2} and~\ref{tab:conv-param-oH2} of Appendix~\ref{app:apdixB}. The values of $j_\mathrm{max}$ here refer to the highest rotational states of the H$_3$O$^+$ cation used in the CC calculations. For the H$_2$ projectile the two lowest rotational states were considered in the scattering calculations, i.e. $j_{\mathrm{H}_2}$ = 0, 2 for {\it p}-H$_2$ and  $j_{\mathrm{H}_2}$ = 1, 3 for {\it o}-H$_2$. The convergence test calculations have shown that the $j_{\mathrm{H}_2}$ = 2 and 3 excited states of H$_2$ have a significant effect on the cross sections' magnitude, and should not be neglected in the CC calculations. For a few energies we performed scattering calculations with a larger H$_2$ rotational basis as well ($j_{\mathrm{H}_2}=0-5$). We found that the impact from these additional hydrogen levels is usually not too large, they have a $\sim 5 \%$ effect on the magnitude of the cross sections in general, which decreases with increasing collision energy. The largest relative difference detected in the collisional data calculated with this extended basis compared to what was computed in the systematic CC calculations is $\sim 25 \%$. Even the $j_{\mathrm{H}_2}=0-3$ basis is very large however, which makes the systematic quantum calculations extremely demanding computationally, especially in the case of the {\it p-}H$_3$O$^+$ target molecule, which is characterised with two times more levels than {\it o-}H$_3$O$^+$ if the same internal energy is considered. Thus, for example, for the {\it p-}H$_3$O$^+$ -- {\it o-}H$_2$ collision in the high energy regime ($\geq 1450$ \cmmo) the number of coupled channels exceeded 8000, for which the computational cost to calculate the cross sections for a defined total energy reaches several thousands of CPU-hours. According to our estimations, about 600.000 CPU-hours were spent in total for the calculation of all H$_3$O$^+$ -- H$_2$ cross sections.

The rate coefficients were computed up to $300$ K kinetic temperatures following the well-known integration method over a Maxwell-Boltzmann distribution of relative velocities:

\begin{equation}
    k_{\mathrm{i} \rightarrow \mathrm{f}}(T) = \left(\frac{8}{\pi\mu k_\mathrm{B}^3 T^3}\right)^\frac{1}{2} \int_{0}^{\infty} \sigma_{\mathrm{i} \rightarrow \mathrm{f}} E_\mathrm{c} e^{-\frac{E_\mathrm{c}}{k_\mathrm{B}T}} dE_\mathrm{c} ,
    \label{eq:rates}
\end{equation}
where $E_\mathrm{c}$ is the collision or kinetic energy, $\sigma_{\mathrm{i} \rightarrow \mathrm{f}}$ is the energy-dependent cross section for the transition from a particular initial (i) to a final state (f), $\mu$ is the reduced mass of the collisional system and $k_\mathrm{B}$ is the Boltzmann constant in atomic units. By summing all possible error contributions (induced both in the PES and scattering calculations) we expect that the accuracy of the collisional rate coefficients is generally as low as a few per cents and always better than $25-30 \%$.

\section{Results and discussion}\label{sec:Results}

In Paper~II we have shown the behaviour of the H$_3$O$^+ -$\textit{p-}H$_2$ rotational de-excitation cross sections and thermal rate coefficients over a limited energy and temperature range. We have presented a comparison of our state-to-state collisional data with the most relevant theoretical works for the H$_3$O$^+ -$ He \citep{ElHanini2019} and for the NH$_3 -$ H$_2$ \footnote{Our motivation for the comparison with ammonia collisional data is related to their isoelectronic nature and structural similarities, and NH$_3$ was assumed to be a template for collisions with H$_3$O$^+$ for a long time.} \citep{Bouhafs2017} collisions. We examined several $\Delta j$ and $\Delta k$ transitions in Paper~II, showing that there is a significant difference between the cross sections we calculated and those of \citet{ElHanini2019} and \citet{Bouhafs2017} in the whole energy range considered (typically our cross sections are about an order of magnitude larger than those from the literature). Another important finding of Paper~II is that there is no direct (linear) scaling between the compared quantities, so the new cross sections and rate coefficients have a significant importance, since they show a large improvement in quality in contrast with the previously published collisional data. Since such a comparative analysis with the available literature was provided and discussed in detail in our previous work, we are not repeating it here.

\subsection{Rotational de-excitation cross sections}\label{sec:CSs}

\begin{figure}
    \centering
		\includegraphics[width=\columnwidth]{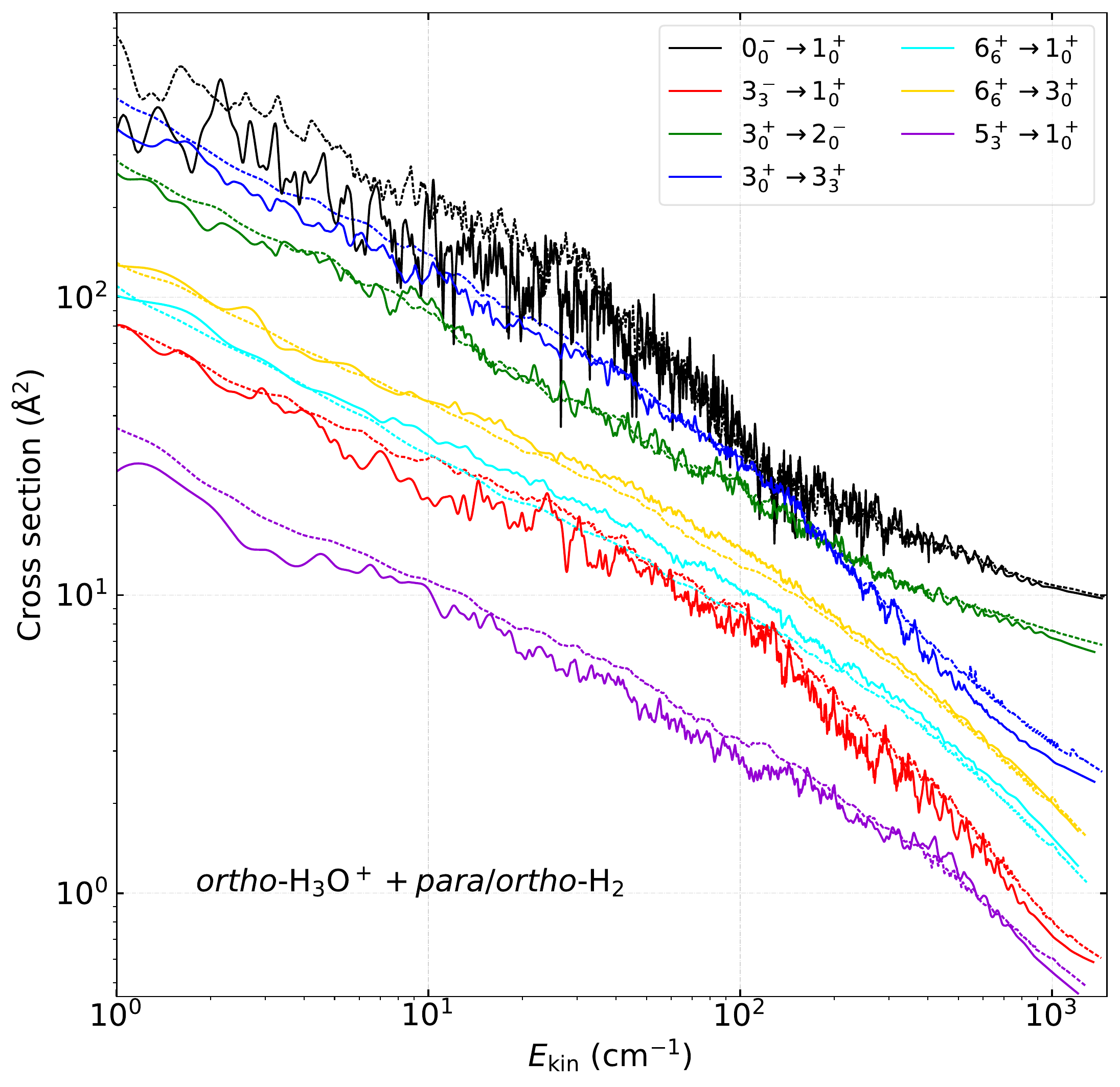}
    \caption{The variation of rotational de-excitation cross sections for some selected transitions of $o-$H$_3$O$^+$ in collision with $p-$H$_2$ (solid lines) and $o-$H$_2$ (dashed lines) projectiles.}
    \label{fig:CSoh3op}
\end{figure}

\begin{figure}
    \centering
		\includegraphics[width=\columnwidth]{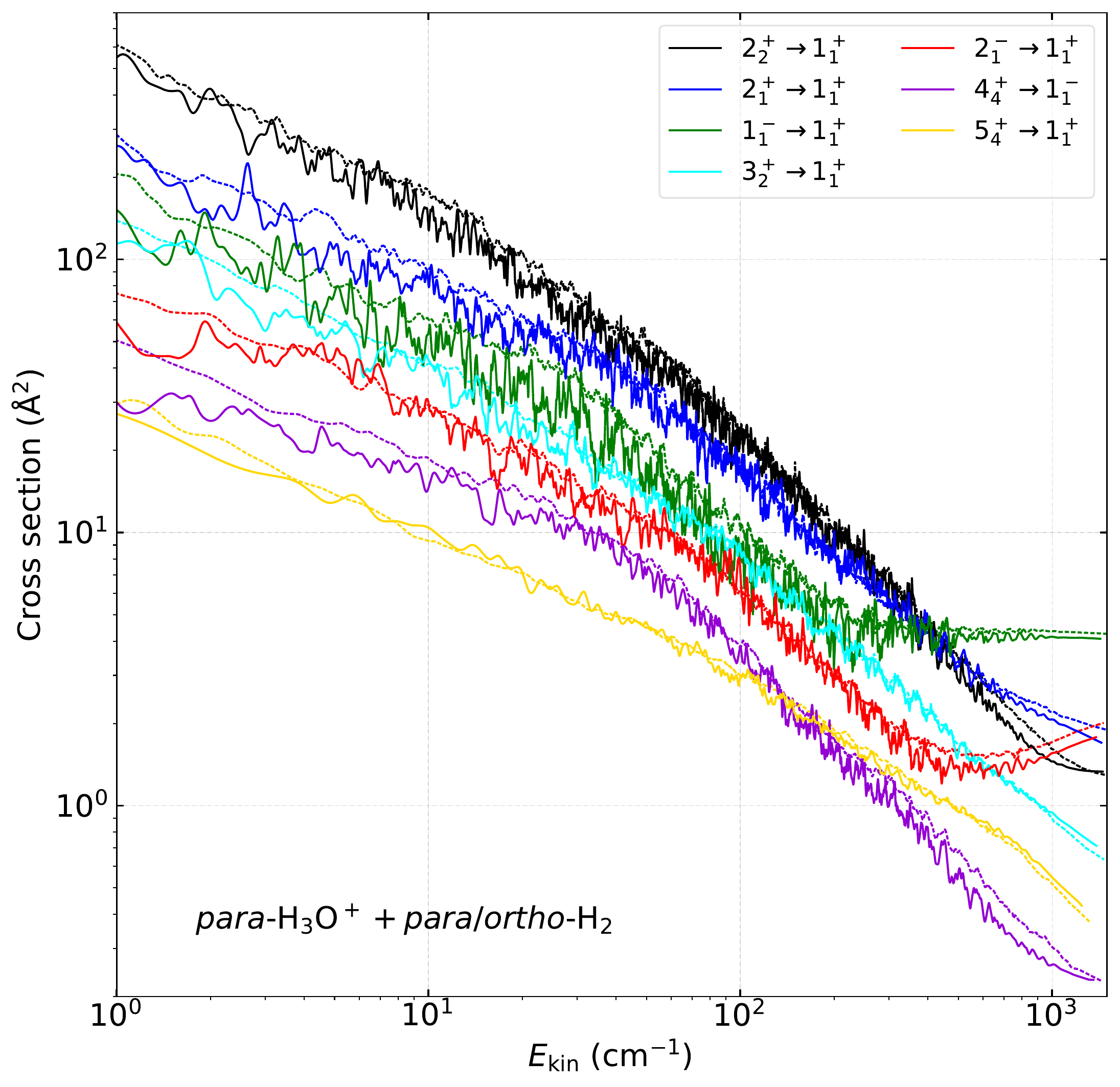}
    \caption{The variation of rotational de-excitation cross sections for some selected transitions of $p-$H$_3$O$^+$ in collision with $p-$H$_2$ (solid lines) and $o-$H$_2$ (dashed lines) projectiles}
    \label{fig:CSph3op}
\end{figure}

In Figs.~\ref{fig:CSoh3op} and~\ref{fig:CSph3op}, the dependence of the H$_3$O$^+$ rotational de-excitation cross sections is presented with respect to the collision (kinetic) energy in collisions with both $p-$H$_2$ (solid lines) and $o-$H$_2$ (dashed lines). In Fig.~\ref{fig:CSoh3op} we compare the cross sections for $o-$H$_3$O$^+$ target species, while Fig.~\ref{fig:CSph3op} shows the corresponding cross sections for $p-$H$_3$O$^+$ for transitions with randomly selected $\Delta j$ and $\Delta k$ parameters and changes in the inversion splitting index.

First, it is important to notice that the cross sections usually monotonically decrease with increasing collision energy (except for some particular transitions) following a typical Langevin-behaviour, as expected for ion-molecule collisions (see, for example, the work by \citet{desrousseaux2019}). The few exceptions (see, for example, the  $2_1^- \rightarrow 1_1^+$ transition in Fig.~\ref{fig:CSph3op}) are most probably related to wide shape resonances, which are associated to quasi-bound states formed due to the multiple secondary wells along the potential energy surface (see Paper~I for PES details). With increasing collision energy (outside the resonances) these cross sections are also expected to follow a Langevin-behaviour, at sufficiently high energies however they will probably rather flatten. Similar cross sections with wide shape resonances were reported earlier for other multi-well collisional systems, like SH$^- - $ He \citep{bop_cold_2017},  C$_5$S $- $ He \citep{khadri_low-temperature_2020}, or  HF $-$ He \citep{stoecklin_vibrational_2003}. The cross sections are characterised by a very dense resonance structure for all transitions, especially at total energies below $800-900$ \cmmo. These Feshbach-type and shape (orbiting) resonances are typical for van der Waals-type complexes with a large well depth. More rigorously, the Feshbach resonances appear due to the formation of bound states via short trapping of the projectile in the potential well, while the shape resonances are related to quasi-bound states which are formed due to tunneling via the centrifugal energy barrier \citep{costes-naulin_2016}. At higher energies the resonances are not as pronounced, so the energy step size was increased.

Our further important finding following the analysis of cross sections is that the data obtained for the $p-$H$_2$ and $o-$H$_2$ projectiles are very similar. This is an expected behaviour, and the same tendency was observed earlier for several other 'ion + molecular hydrogen' collision schemes, for example recently for NS$^+ +$ H$_2$ \citep{bop2022}, HCS$^+ +$ H$_2$ \citep{denis-alpizar2022}, HCO$^+$ and DCO$^+ +$ H$_2$ \citep{denis-alpizar2020}, CF$^+ +$ H$_2$ \citep{desrousseaux2019} and also for collisions of negative ions like C$_3$N$^- +$ H$_2$ \citep{lara-moreno2019} or CN$^- +$ H$_2$ \citep{klos-lique2011}. According to \citet{lara-moreno2019} these similarities between the $p-$H$_2$ and $o-$H$_2$ cross sections could be attributed to the features of the short-range interaction of the colliders, which gives relevant contributions in the coupling matrix elements equally for $p-$H$_2$ and $o-$H$_2$. Also, the long range part of the PES is weakly anisotropic with respect to the rotation of H$_2$ (see Paper~I for details).

In accordance with the results of previous authors, the cross sections we computed with the {\it ortho}-H$_2$ projectile are somewhat larger compared to the {\it para}-H$_2$ data, but the differences usually do not exceed $10 \%$. It is worth noticing that for some transitions we observe more pronounced differences (up to about $20-30 \%$) at higher energies ($\geq 400$ \cmmo), which is different from what was generally found for other ion + molecule collisions, where the main differences were present in the low-energy regime (see, for example, the works by \citet{bop2022} and \citet{denis-alpizar2022}). This feature is most probably associated to the larger anisotropy of the PES with respect to H$_2$ rotation in the short range. The resonances are not so remarkable in the case of the $o-$H$_2$ collider (analogously as it was found by other systems), but this feature is likely to be related to the high number of overlapping resonances due to the significantly higher number of channels and couplings compared to $p-$H$_2$ (remember, the rotational basis involves the $j_{\mathrm{H}_2}$ = 0, 2 for $p$-H$_2$ and  $j_{\mathrm{H}_2}$ = 1, 3 states for $o$-H$_2$).

\subsection{Rate coefficients}\label{sec:RCs}

\begin{figure}
    \centering
		\includegraphics[width=\columnwidth]{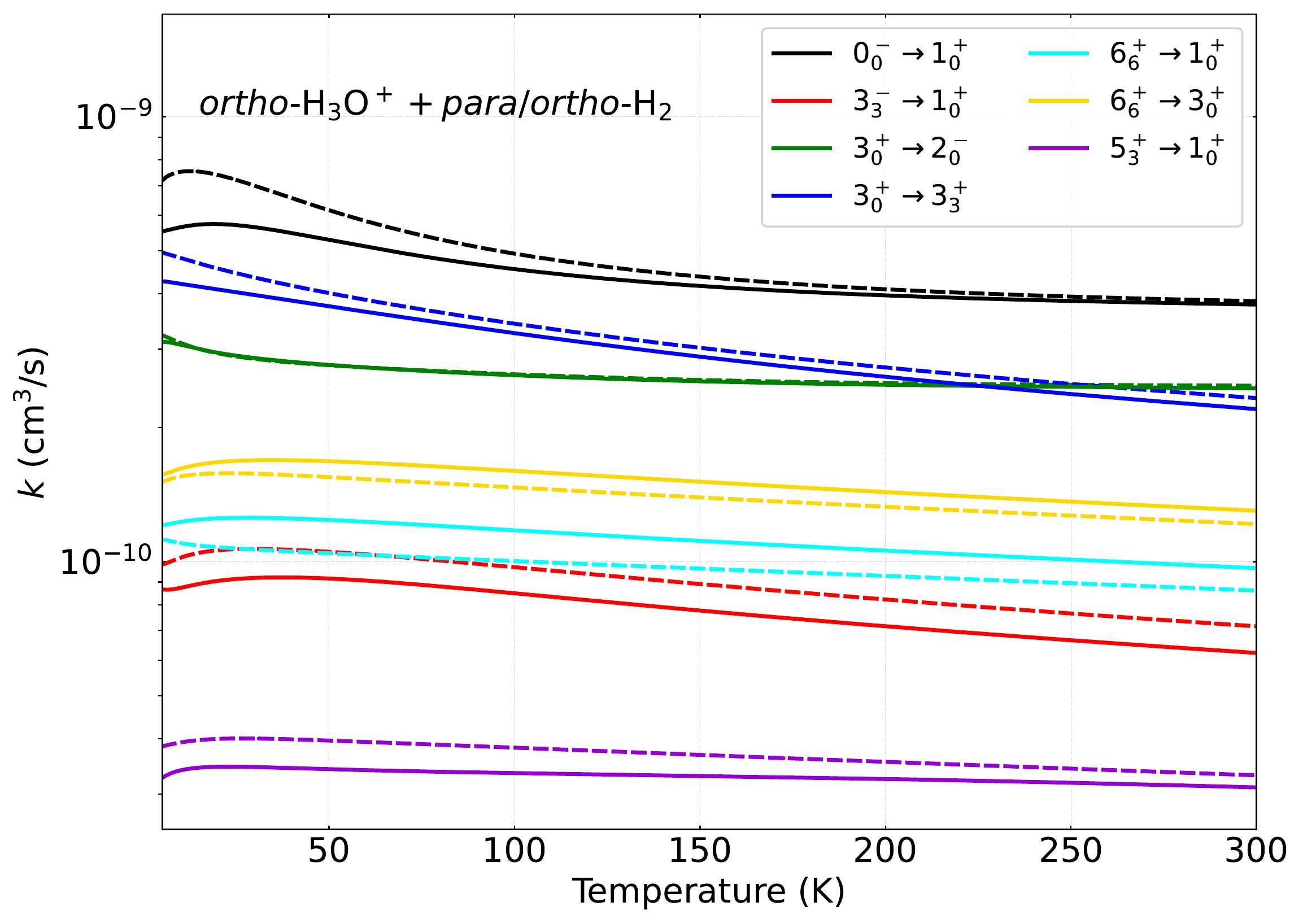}
    \caption{Kinetic temperature dependence of the rate coefficients for the same rotational transitions as shown in  Fig.~\ref{fig:CSoh3op} for $o-$H$_3$O$^+$ collision with $p-$H$_2$ (solid lines) and $o-$H$_2$ (dashed lines) projectiles.}
    \label{fig:RCoh3op}
\end{figure}

Once the state-to-state cross sections were calculated we computed the corresponding rate coefficients for the H$_3$O$^+$ -- H$_2$ collision (see subsection~\ref{sec:ScatCalc} for details). In Fig.~\ref{fig:RCoh3op} the rate coefficients for the $o-$H$_3$O$^+ -$ collision with both $p$-H$_2$ and $o$-H$_2$ colliders are compared for the same rotational transitions which are presented in Fig.~\ref{fig:CSoh3op}. In the whole kinetic temperature interval from 10 to 300 K the temperature dependence is relatively weak, which is in accordance with the Langevin capture model. Analogously as in the case of the cross sections, the rate coefficients calculated for $p$-H$_2$ and $o$-H$_2$ are very similar both in their behaviour and magnitudes, in agreement with previous findings in the literature (see, for example, \citet{bop2022, denis-alpizar2022,desrousseaux2019,klos-lique2011}). The magnitude of the rate coefficients with $o$-H$_2$ is somewhat higher for all transitions. These differences are almost constant above 100 K, and usually do not exceed $10 \%$ except for some particular transitions for which they can reach a maximum of $15-20 \%$.

It is worth noting also that there is no constant linear scaling (uniform temperature dependence) observed between the rate coefficients compared in Fig.~\ref{fig:RCoh3op} with respect to the change in kinetic temperature. If we examine for instance the $0_0^- \rightarrow 1_0^+$ and $3_0^+ \rightarrow 2_0^-$ transitions, we can see some significant differences: for collision with $o$-H$_2$ the rate coefficients for the former decrease from $7.5 \cdot 10^{-9}$ down to $3.8 \cdot 10^{-9}$ cm$^3$s$^{-1}$ (i.e. by a factor of 2) between 15 and 300 K, while for the latter transition the rate coefficients are quasi constant in the whole temperature range. The rate coefficients exhibit a non-linear behaviour on the logarithmic scale dominantly only in the low-temperature regime, i.e. below 50 K. The close magnitude and similar temperature dependence of the state-to-state rate coefficients computed with $p$-H$_2 (j=0)$ and $o$-H$_2 (j=1)$ colliders imply a weak dependence of the collisional data on the rotational state of H$_2$. Based on this one can expect that the rate coefficients for H$_3$O$^+$ are very similar in collision with a rotationally hot hydrogen projectile ($j_{\mathrm{H}_2} = 2,3,...$)

\begin{figure}
    \centering
		\includegraphics[width=\columnwidth]{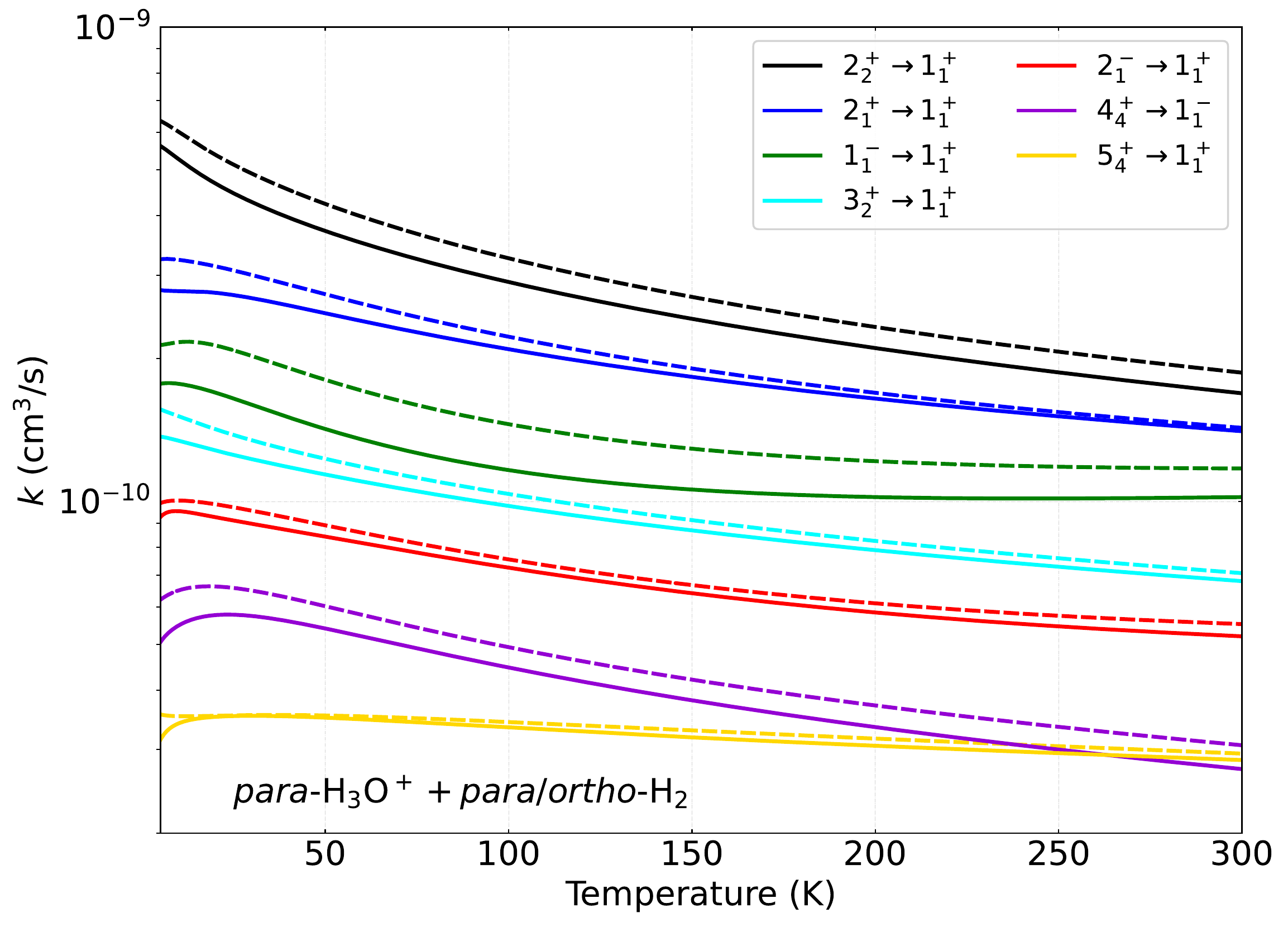}
    \caption{Kinetic temperature dependence of the rate coefficients for the same rotational transitions as shown in  Fig.~\ref{fig:CSph3op} for $p-$H$_3$O$^+$ collision with $p-$H$_2$ (solid lines) and $o-$H$_2$ (dashed lines) projectiles.}
    \label{fig:RCph3op}
\end{figure}

In Fig.~\ref{fig:RCph3op} the rate coefficients for the $para-$H$_3$O$^+$ collision with both $para$-H$_2$ and $ortho$-H$_2$ nuclear spin isomers are compared again up to $300$ K for the same rotational transitions which are presented in Fig.~\ref{fig:CSph3op}. The overall behaviour and the variation of the rate coefficients with temperature are very similar to what was found in the case of the $ortho-$H$_3$O$^+$ nuclear spin species. The rate coefficients computed with $p$-H$_2$ and $o$-H$_2$ colliders are similar, the differences between them is similar to those discussed in the paragraph above.

We have compared our rate coefficients with the corresponding data from the \texttt{LAMDA} data base (\citet{Schoier2005}, see also \citet{VanDerTak2020}). Currently these latter are used for astrophysical modelling of hydronium in interstellar clouds, but the data are rather limited: only a single set of rate coefficients are provided for H$_3$O$^+$ for 100 K kinetic temperature, and only with $para-$H$_2$ collider. Another important limitation is that the data base involves collisional and spectroscopic data for states with $K \leq 3$ only, missing some important states which can be populated $\sim 100$ K. The collisional rate coefficients in the current LAMDA datafile were estimated from scaled radiative rates, following the scaling procedure described by \citet{VanDerTak2020}. For the non-radiative (dipole-forbidden) transitions, NH$_3$ collisional data of \citet{danby88} were adopted, because NH$_3$ is isoelectronic with H$_3$O$^+$ and the two species are very similar in molecular mass. This places the H$_3$O$^+$ \texttt{LAMDA} datafile in the category "rough scalings / similar species", which has an estimated uncertainty of a factor of 10, according to \citet{VanDerTak2020}. In contrast to this, we are providing exact rate coefficients for all nuclear spin species (including $ortho-$H$_2$) for a wide range of $T_\mathrm{kin}$ from 10 up to 300 K.

\begin{figure}
    \centering
        \includegraphics[width=\columnwidth]{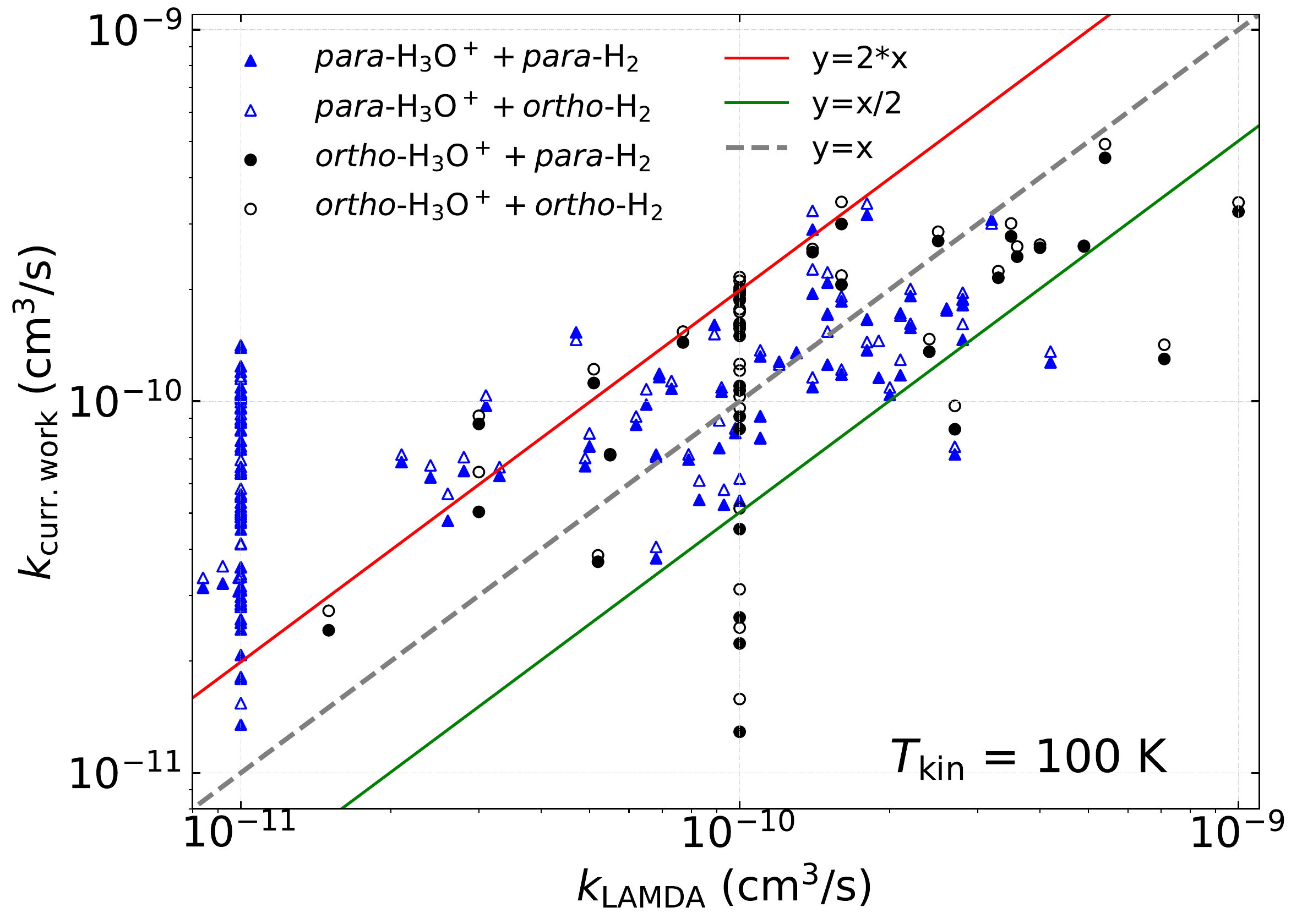}
    \caption{Ratios of our recent state-to-state thermal rate coefficients for all nuclear spin configurations of the H$_3$O$^+ +$ H$_2$ collision with respect to the corresponding data listed in the \texttt{LAMDA} data base \citep{Schoier2005} at 100~K kinetic temperature. The reference collisional data \texttt{LAMDA} data base are only available for $p-$H$_2$ collider.}
    \label{fig:RCcomp-sts}
\end{figure}

Due to the mentioned limitations of the reference data, in Fig.~\ref{fig:RCcomp-sts} we present ratios of the state-to-state rate coefficients with respect to the corresponding data from \texttt{LAMDA} at 100 K, for both $o-$H$_3$O$^+$ and $p-$H$_3$O$^+$. As one can see, the ratios show a large scatter: no linear scaling trends can be observed between our collisional rate coefficients and those from the \texttt{LAMDA} data base. The calculated ratios imply significant differences between the compared sets of rate coefficients, with typical differences of up to a factor of 2 in both directions. Due to the rough scaling, some of the rate coefficients in the \texttt{LAMDA} data base are likely to be overestimated at values of $10^{-10}$ cm$^3$s$^{-1}$, while on the opposite side, some lower-magnitude collisional data were likely underestimated (in particular the series of rate coefficients at $10^{-11}$ cm$^3$s$^{-1}$). Therefore it is not surprising that our quantum-level calculations show differences even up to a factor of 10 with the roughly scaled data.

Compared to the large variations with respect to the \texttt{LAMDA} values, no significant differences can be found in Fig.~\ref{fig:RCcomp-sts} between our data computed with $o/p-$H$_2$ colliders. Some slight tendencies could be observed however. For instance, the large-magnitude rate coefficients from \texttt{LAMDA} systematically overestimate our collisional data. For the low-magnitude transitions (with values below about $2 \times 10^{-10}$ cm$^3$s$^{-1}$) we find an opposite trend however: the \texttt{LAMDA} data are significantly lower than ours. This means that the collisional data, which were used for astrophysical modelling before, overestimated the de-excitation probability of the dominant transition channels, while the contributions from weaker channels were underestimated. All these could lead to large uncertainties in previous modellings, and the improved set of rate coefficients we present here will have a significant impact on the abundance and excitation modelling of hydronium, especially in warm dense molecular clouds in the ISM.

\section{Radiative transfer modelling of hydronium in warm interstellar clouds}\label{sec:Excit}

Radiative transfer modelling was performed to study the excitation of H$_3$O$^+$ in astrophysical media based on the new set of rate coefficients with \textit{ortho-} and \textit{para-}H$_2$ colliders. The calculations were performed using the latest version of the \texttt{RADEX} non-LTE radiative transfer computer program \citep{VanDerTak2007} based on the large velocity gradient (LVG) approximation. The new set of rate coefficients allows one to explore a significantly larger range of kinetic temperatures (up to $T = 300$ K) compared to our previous work (Paper~II) and also supports more reliable radiative transfer modelling for these high temperatures due to the new, previously missing collisional data for \textit{o-}H$_2$, which becomes the dominant collider above 100 K. The $ortho$-to-$para$ ratio of H$_2$ at different temperatures was considered based on an LTE-distribution, as implemented in the \texttt{RADEX} software: $0.23/0.77$ at $50$ K, $0.62/0.38$ at $100$ K, $0.75/0.25$ at $200$ and $300$ K. However, due to the similarities between the $o/p$-H$_2$ collisional data we do not expect significant impact on the modelling with respect to the variation of the $ortho$-to-$para$ ratio.

The aim of the radiative transfer modellings performed in present work is to demonstrate the impact of our new rate coefficients on the inferred column densities, and not to re-analyse and re-interpret the previous observations. For this purpose we compare the radiation (brightness) temperatures computed by using two different sets of collisional rate coefficients (similar modelings were performed in Paper~II). For the first (reference) set we adopted the available rate coefficients for H$_3$O$^+$ from the \texttt{LAMDA} data base \citep{Schoier2005}, which are available only for at 100 K and were derived from scaled NH$_3$ + \textit{p-}H$_2$($j=0$) collisional data of \citet{danby88}, considering also the cross sections reported by \citet{Offer1992}. As the second set we have used our new rate coefficients, which cover a much larger temperature range ($10-300$ K) and all possible nuclear species, including \textit{o-}H$_2$. The rotational energies along with the frequencies and Einstein $A$-coefficients of the modelled radiative transition were taken from the JPL data base (\citet{Pickett2010}, Species Tag: 19004, version 3, compiled by Yu \& Drouin, Jan. 2010). For more details about the states involved in the calculations, see Table.~1 in Paper~II.

As shown in Section~\ref{sec:ObsLit}, H$_3$O$^+$ was detected in several environments, which possess very different physical conditions. For example, in diffuse molecular clouds H$_2$-densities from $35$ up to $10^3$ \pcmc were reported, while for dense clouds the hydrogen densities can be as high as $10^8$ \pcmc, but are typically around $10^6$ \pcmc. For column densities the observations report several values in a broad range from $0.7 \times 10^{13}$ up to $7 \times 10^{16}$ \pcm. The typical temperatures of the environments, where hydronium was detected, are $ \geq 100$ K, but some authors reported higher temperatures as well ($\sim200, \sim240, \sim380$ and even $\sim500$ K). In diffuse ISM sources H$_3$O$^+$ lines are usually detected in absorption \citep{Gerin2010,lis14, Indriolo2015}. It is worth mentioning however that emission from the $0_0^- \rightarrow 1_0^+$ (984.7 GHz) line is also reported by \citet{Indriolo2015}, which confirms the necessity of collisional rate coefficients even for low-density clouds. For this reason we decided to calculate the radiation temperatures for all observed transitions H$_3$O$^+$ under specific constrained physical conditions, which were carefully selected based on the reported observations. In particular, $50, 100, 200$ and $300$ K kinetic temperatures were considered to cover the majority of the observations, but not going beyond the upper limit of our rate coefficients. We separately modelled diffuse and dense cloud conditions with hydrogen densities of $10^3$ and $10^6$ \pcmc, respectively. For column densities we have decided to sample strictly optically thin regions with $1 \times 10^{13}$ \pcm as well as optically thick clouds with $1 \times 10^{16}$ \pcm, which nearly cover the lower and upper boundaries of the reported observations. For the modellings in the optically thin region the opacity $\tau \ll 1$, while under optically thick conditions $\tau \simeq 1-1000 $ except for some particular transitions at low H$_2$-density ($10^3$ \pcmc). The background temperature and the line width parameters were kept fixed in the calculations at $T_\mathrm{bg} = 2.7$~K and $\Delta V = 1.0$~km/s, respectively.

\begin{figure*}
		\includegraphics[width=0.9\textwidth,center]{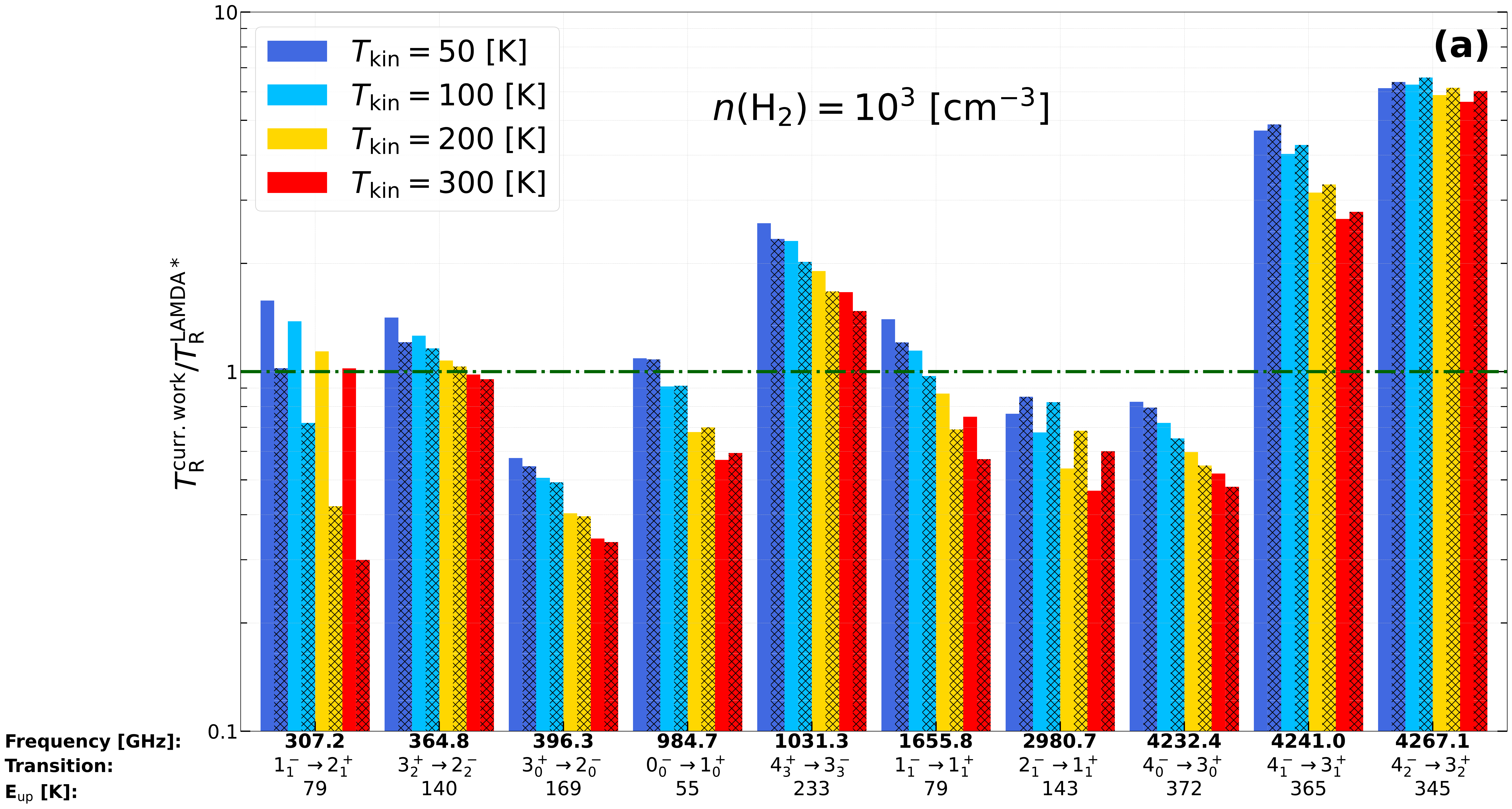}
    \label{fig:TR-pH3Op-a}

		\includegraphics[width=0.9\textwidth,center]{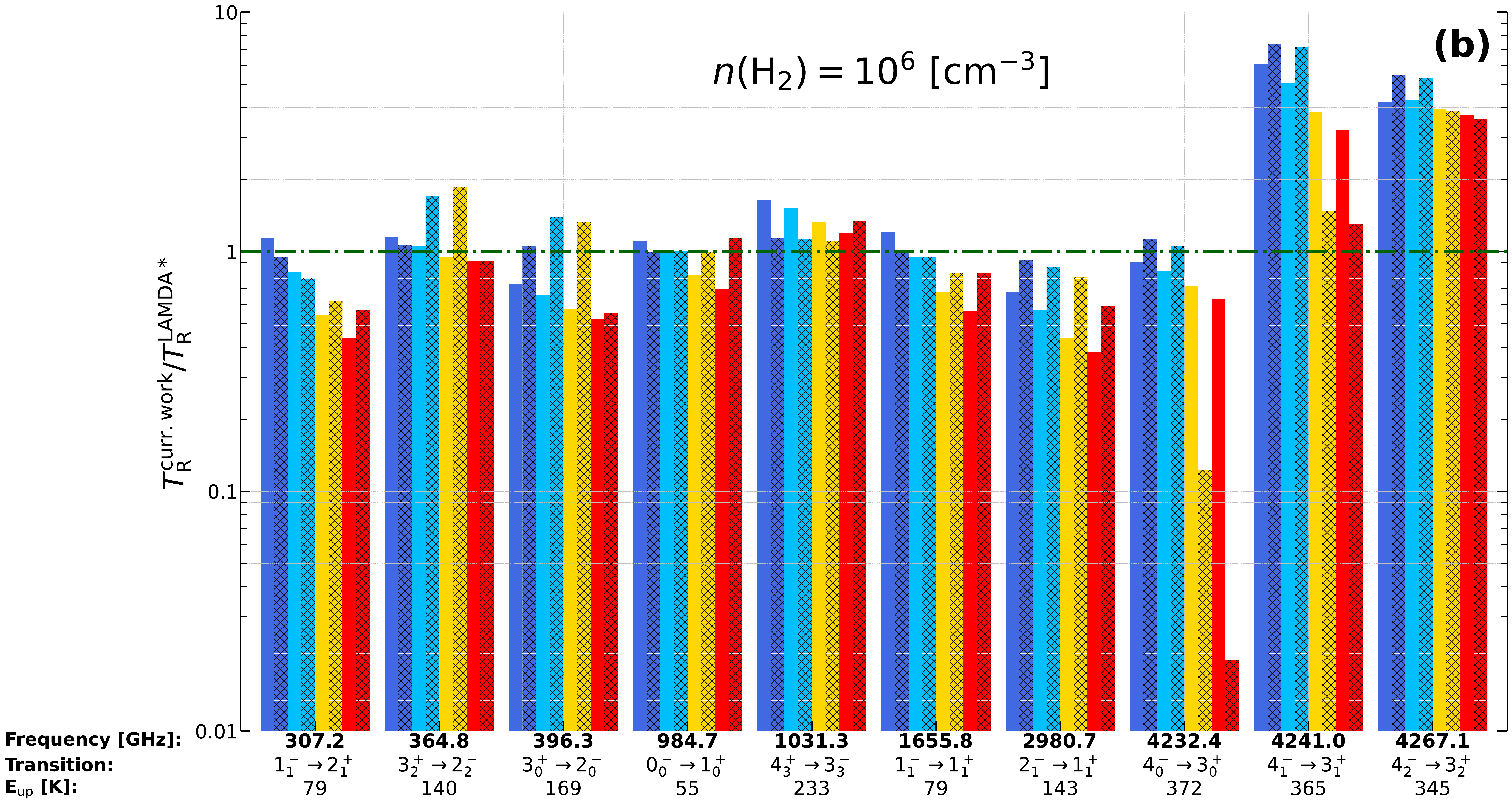}
    \caption{The relative ratio of radiation temperatures at 50, 100, 200 and 300 K kinetic temperatures for the majority of the observed radiative H$_3$O$^+$ transitions calculated for diffuse (panel (a), $n(\mathrm{H}_2) = 10^3$ \pcmc) and dense (panel (b), $n(\mathrm{H}_2) = 10^6$ \pcmc) molecular cloud conditions. The $T_\mathrm{R}^\mathrm{curr.work}$ data is computed from our new set of rate coefficients, while the $T_\mathrm{R}^\mathrm{LAMDA^*}$ set is calculated based on the corresponding collisional data from the \texttt{LAMDA} data base \citep{Schoier2005}. The colour bars correspond for radiative transfer calculations in the optically thin environments  ($N = 10^{13}$ \pcm), while the hatch-filled bars show the same quantities, but for optically thick regions ($N = 10^{16}$ \pcm).}
    \label{fig:TR-pH3Op-b}
\end{figure*}

To demonstrate the reliability of our modellings, in Paper~II we have calculated the radiation temperatures at $T_\mathrm{kin}=50$ and $100$ K for the $1_1^- \rightarrow 2_1^+$, $3_2^+ \rightarrow 2_2^-$ and $3_0^+ \rightarrow 2_0^-$ transitions, which showed a good quantitative agreement with the results of \citet{Phillips1992}. In this paper however, we compare the relative ratio of radiation temperatures for the observed radiative transitions of both {\it ortho-} and {\it para-}H$_3$O$^+$ calculated for diffuse (Fig.~\ref{fig:TR-pH3Op-b}.(a)) and dense (Fig.~\ref{fig:TR-pH3Op-b}.(b)) molecular cloud conditions. The $T_\mathrm{R}^\mathrm{curr.work}$ radiation temperatures are computed from our new set of rate coefficients, while the $T_\mathrm{R}^\mathrm{LAMDA^*}$ set is calculated based on the corresponding collisional data from the \texttt{LAMDA} data base \citep{Schoier2005}. The solid colour bars correspond to radiative transfer calculations in the optically thin environment ($N = 10^{13}$ \pcm), while the hatch-filled colour bars show the same ratios at large column densities ($N = 10^{16}$ \pcm).

As one can see in Fig.~\ref{fig:TR-pH3Op-b}, there are very significant differences between the radiation temperatures calculated with the two set of rate coefficients. A ratio close to unity is observed only for very few transitions, under specific physical conditions. Also, no coherence patterns were found for the data presented, and the $T_\mathrm{R}$ ratios depend very strongly both on the kinetic temperature, on the H$_3$O$^+$ column and H$_2$ volume densities as well as on the particular rotational transition. There are some slight trends towards temperature dependence only, since the ratios are usually decreasing as $T_\mathrm{kin}$ increases. It is not a strict rule however and several deviations from this are observed, most notably for larger column densities (i.e. for $10^{16}$ \pcm). Moreover, since this tendency is mostly valid for the $T_\mathrm{R}$ ratios much below $1$ as well, we can observe the highest relative deviations at higher  $T_\mathrm{kin}$. In particular, the radiation temperature ratio can be as low as $\sim 0.02$ for the $4_0^- \rightarrow 3_0^+$ transition at 4232 GHz in Fig.~\ref{fig:TR-pH3Op-b}.(b), which means that the $T_\mathrm{R}$ calculated for this line with our new set of rate coefficients is more than a factor of 50 smaller than that is calculated from the \texttt{LAMDA} data. On the other hand, for the very close $4_1^- \rightarrow 3_1^+$  {\it para-}H$_3$O$^+$ transition at 4241 GHz an opposite tendency is observed: at 50 and 100 K differences up to more than a factor of 7 are found in the other direction. For most transitions the $T_\mathrm{R}$ ratios vary between $0.5$ and $2$. Larger than a factor of 3 differences are found for the $4_1^- \rightarrow 3_1^+$ and $4_2^- \rightarrow 3_2^+$ transitions most notably, as well as for some other transitions at 300 K (see the 307 and 396 GHz lines at $10^3$ \pcmc and the 2981 and 4232 GHz lines at $10^6$ \pcmc gas density). The observed large variation of the ratios are not only associated with the differences in the magnitude of the corresponding rate coefficients, but also to the larger temperature range covered, larger number of rotational states involved as well as to the presence of collisional data with {\it ortho-}H$_2$ collider presented in this work (although the impact is limited to $ \sim 20\%$). These conclusions are supported also by the fact that large differences are found also for radiation temperatures, which are calculated close to local thermodynamic equilibrium conditions (with $n(\mathrm{H}_2) = 10^6$ \pcmc), where the state-to-state collisional rate coefficients usually have less influence on the results of radiative transfer modellings. About the rotational states involved it is important to mention that we have considered 21 {\it para-} and 11 {\it ortho-}H$_3$O$^+$ levels, including states with $K \leq 6$ quantum numbers, while the \texttt{LAMDA} data base provides rate coefficients for 14 and 9 levels of them, respectively, with $K \leq 3$ only. In addition to the higher rotational states (i.e. those with $J = 5$), the \texttt{LAMDA} set of rate coefficients also excludes some of the lower states (see Table~\ref{tab:rot-levels} in Appendix~\ref{app:apdixA} for details), which can be populated already at moderate temperatures (from about 100 K). Due to this, some of the rotational levels could be overpopulated in radiative transfer modelling, which can lead to additional significant differences when using the \texttt{LAMDA} rate coefficients or those from this work. To estimate the magnitude of these differences, we have also computed the $T_\mathrm{R}$ ratios with an equivalent set of rotational states, thus involving our present rate coefficients only for those levels, which are present in the \texttt{LAMDA} data base, and neglecting all others in the radiative transfer model. We found that the ratios of the radiation temperatures still significantly deviate from unity, which confirms that the largest impact is related to the differences in magnitude of the rate coefficients. The other, non-negligible, contribution is due to the missing levels and transitions. It is also important to notice that population inversion can be observed for several transitions of hydronium under non-LTE conditions, and the new collisional data have a significant impact of their characteristics even at lower kinetic temperatures, as we demonstrated earlier in Paper~II for some low-frequency transitions at 50 and 100 K (see subsection 3.3 therein).

In general, Fig.~\ref{fig:TR-pH3Op-b}.(a) and (b) show that the variation of the $T_\mathrm{R}^\mathrm{curr.work} /  T_\mathrm{R}^\mathrm{LAMDA^*}$ relative ratios are large and not clearly correlated to any physical properties. The ratios are changing notably from transition to transition and there are no unambiguous patterns of their dependence on the particular upper or lower rotational states. There is no correlation found neither with respect to molecular, neither column densities. The only systematic behaviour found is the slight kinetic temperature dependence, which is also not valid for all transitions and at any density conditions. Based on these findings we conclude that the new rate coefficients proposed in present work have a significant impact on astrophysical modellings, more particularly on the determination of column densities. This can lead consequently to more precise interpretation of H$_3$O$^+$ observations in various dense and diffuse interstellar environments by deriving more accurate hydronium abundances, which can be remarkably different from those reported previously for this species.

\section{Conclusions}\label{sec:Conclusions}

We have presented in this work state-to-state rotational de-excitation cross sections and thermal rate coefficients for the collision of hydronium (H$_3$O$^+$) with molecular hydrogen. For theoretical calculations the best available and most accurate close-coupling scattering theory has been used based on our state-of-the-art five-dimensional rigid-rotor potential energy surface (Paper~I). All nuclear spin configurations ({\it ortho/para}) are considered for both colliding partners, so the collisions with {\it ortho-}H$_2$ were studied as well, which is the most abundant collider in interstellar clouds with temperatures above 100 K.

As it is expected to be valid for all ion + molecule collisional systems, most of the cross sections monotonically decrease as the collision energy is increasing, following a typical Langevin-behaviour. They  exhibit a very dense resonance behaviour, especially below $800-900$ \cmmo kinetic energies, which are associated with the formation of bound and quasi-bound states in the potential well. We have also found that there are no significant differences between the collisional data computed with {\it ortho-} and {\it para-}H$_2$ colliders. Due to this one can expect that H$_3$O$^+$ collisional data are similar for collisions with rotationally hot hydrogen ($j_{\mathrm{H}_2} = 2,3,...$). We have observed also a relatively weak temperature dependence in the case of the rate coefficients.

Having a full set of collisional data for all rotational levels of hydronium below $420$ K ($292$ cm$^{-1}$), we have presented a comparison for all particular rotational transitions individually with the available rate coefficients from the \texttt{LAMDA} data base \citep{Schoier2005}, which were used earlier by many authors for the interpretation of H$_3$O$^+$ observations in interstellar clouds. Significant differences are found between the two data sets, typically up to a factor of 2, but occasionally up to a factor of 10. The tendencies we observed in this analysis indicate that the collisional data, which were used earlier for astrophysical modelling, overestimate the de-excitation probability of the dominant channels, while the weaker channels are underestimated.

In order to estimate the impact of the new rate coefficients on the astrophysical models, we have performed radiative transfer modelling of H$_3$O$^+$ based on the the LVG approximation by the \texttt{RADEX} non-LTE code. We compare the radiation temperatures computed by using two sets of rate coefficients: first, the available data for H$_3$O$^+$ from the \texttt{LAMDA} data base \citep{Schoier2005} and, second, the new collisional rate coefficients presented in this work. We observed very significant differences between the $T_\mathrm{R}$ calculated with the two set of rates, once we compared their relative ratios. For most of the lines the ratios of the calculated radiation temperatures are far from unity, typically within factors of 2 in either direction, which is similar to what was found for the rate coefficients. The ratios also depend very strongly on the kinetic temperature, column and hydrogen densities as well as on the particular transitions, so they do not correlate with the physical conditions. It is important to notice that these large variations of the ratios are not only associated with the differences between the state-to-state rate coefficients, but also with the larger temperature range covered, the larger number of rotational levels involved as well as to the additional collisional data for {\it ortho-}H$_2$.

Based on the results of the analysis in the present work we conclude that the new, accurate (within $10-20 \%$) rate coefficients without doubt have a significant impact on the radiative transfer modelling of astronomical environments, including dense and diffuse molecular clouds. Consequently, the new collisional data allow a more adequate interpretation of hydronium observations in interstellar clouds, leading to more accurate estimations for column densities and relative abundances of hydronium, especially in warmer molecular clouds of the ISM at $T \geq 100$ K. This ensures a more robust indirect way to estimate the rate of O$_2$ and H$_2$O production in interstellar regions.

\section*{Acknowledgements}

We acknowledge financial support from the European Research Council (Consolidator Grant COLLEXISM, Grant Agreement No. 811363) and the Programme National “Physique et Chimie du Milieu Interstellaire” (PCMI) of CNRS/INSU with INC/INP cofunded by CEA and CNES. We wish to acknowledge the support from the CEA/GENCI for awarding us access to the TGCC/IRENE supercomputer within the A0110413001 project and also the KIF\"{U} for awarding us access to HPC resources based in Hungary. S.D. acknowledges the support from COST Action CA18212 - Molecular Dynamics in the GAS phase (MD-GAS), supported by COST (European Cooperation in Science and Technology). F.L. acknowledges the Institut Universitaire de France.

\vspace{12pt}
{\it The authors dedicate this paper to the memory of Tom Phillips, a pioneer of submillimeter observations in general and astronomical H$_3$O$^+$ spectroscopy in particular, who passed away on August 6, 2022, at the age of 85.}

\section*{DATA AVAILABILITY}

The data that support the findings of this study are available within the article and its supplementary material at MNRAS online.

The molecular data files in \texttt{RADEX}-format, which were used in the radiative transfer calculations, and which include the particular state-to-state rate coefficients for {\it ortho/para-}H$_3$O$^+$ collisions with {\it ortho/para-}H$_2$ up to 300 K are provided as supplementary material. The data files containing the energy-dependent de-excitation cross sections are also provided as supplementary material.

The data underlying this article will be made publicly available also through the \texttt{EMAA} \url{https://emaa.osug.fr/}, \texttt{LAMDA} \url{https://home.strw.leidenuniv.nl/~moldata/} and \texttt{BASECOL} \url{https://basecol.vamdc.eu/} data bases.




\bibliographystyle{mnras}
\bibliography{references}



\appendix

\section{A note on the rotational levels of H$_3$O$^+$ considered in the scattering calculation}\label{app:apdixA}

The complete set of the rotational states $j_k^{\epsilon}$ considered in this work both for {\it ortho-} and {\it para-}H$_3$O$^+$ are provided in Table~\ref{tab:rot-levels} ($j$ is the total angular momentum of the H$_3$O$^+$ cation, $k$ is its projection on the $C_3$ rotational axis, and $\epsilon=\pm$ is an inversion symmetry index). {\it Ortho-}H$_3$O$^+$ is characterised with $k=3n$ quantum numbers (where $n=0,1,2,\dots$), while all other $k$ quantum numbers (e.g. $k=1,2,4,5,\dots$) refer to {\it para-}H$_3$O$^+$. For {\it ortho}-H$_2$ and {\it para}-H$_2$ the two lowest even rotational states were taken into account in the scattering calculations, i.e. $j_{\mathrm{H}_2} = 0, 2$ and  $j_{\mathrm{H}_2} = 1, 3$, respectively.

\begin{table*}
	\centering
    \caption{List of the rotational energy levels (in \cmmo and K) for {\it ortho-} and {\it para-}H$_3$O$^+$, which were considered in collisional studies of the present work. The corresponding energies are taken from JPL data base (\citet{Pickett2010}, Species Tag: 19004, version 3, compiled by Yu \& Drouin, Jan. 2010). The state labels with star (*) symbol indicate those H$_3$O$^+$ levels, which are missing in the \texttt{LAMDA} data base \citep{Schoier2005}.}
    \label{tab:rot-levels}
    \begin{tabular}{cccc|cccc}
        \hline
        \multicolumn{4}{c}{\textbf{\textit{para-}H$_3$O$^+$}} &  \multicolumn{4}{c}{\textbf{\textit{ortho-}H$_3$O$^+$}}\\
        \hline
        state  & rotational  & \multicolumn{2}{c}{rotational energy} & state  & rotational  & \multicolumn{2}{c}{rotational energy} \\
        label & state $j_k^{\epsilon}$  & [\cmmo] & [K] & label & state $j_k^{\epsilon}$ & [\cmmo] & [K] \\
        \hline
        (1) & $1_1^+$	&   0.000 &  0.0    & (1) & $1_0^+$ &   5.101	&   7.3 \\
        (2) & $2_2^+$	&  29.697 &  42.4   & (2) & $0_0^-$ &  37.947	&   54.2    \\
        (3) & $2_1^+$	&  44.986 &  64.3   & (3) & $3_3^+$ &  71.681	&   102.4   \\
        (4) & $1_1^-$	&  55.233 &  78.9   & (4) & $2_0^-$	& 104.239	&   148.9   \\
        (5) & $2_2^-$	&  84.977 &  121.4  & (5) & $3_0^+$	& 117.457	&   167.8   \\
        (6) & $3_2^+$	&  97.145 &  138.8  & (6) & $3_3^-$	& 127.172	&   181.7   \\
        (7) & $2_1^-$	&  99.427 &  142.1  & (7) & $4_3^+$ & 161.573	&   230.9   \\
        (8) & $3_1^+$	& 112.384 &  160.6  & (8) & $4_3^-$	& 215.486	&   307.9   \\
        (9)* & $4_4^+$	& 125.939 &  179.9  & (9) & $4_0^-$	& 258.636	&   369.5   \\
        (10) & $3_2^-$	& 151.241 &  216.1  & (10)* & $6_6^+$ & 271.201	&   387.5   \\
        (11) & $3_1^-$	& 165.657 &  236.7  & (11)* & $5_3^+$ & 273.697	&   391.1   \\
        (12)* & $4_4^-$	& 181.807 &  259.8  \\
        (13) & $4_2^+$	& 186.926 &  267.1  \\
        (14)* & $5_5^+$	& 192.453 &  275.0  \\
        (15) & $4_1^+$	& 202.099 &  288.8  \\
        (16)* & $5_4^+$	& 238.256 &  340.4  \\
        (17) & $4_2^-$	& 239.479 &  342.2  \\
        (18)* & $5_5^-$	& 248.863 &  355.6  \\
        (19) & $4_1^-$	& 253.850 &  362.7  \\
        (20)* & $5_4^-$	& 292.146 &  417.4  \\
        (21)* & $5_2^+$	& 298.914 &  427.1  \\ 
        \hline
    \end{tabular}
\end{table*}

\section{A note on the converged values of the basis set size and total angular momenta in scattering calculations}\label{app:apdixB}

The values of the $j_\mathrm{max}$, $J_\mathrm{tot}$ and $E_\mathrm{step}$ parameters are discussed here for all total energy intervals ranging from $E_\mathrm{init}$ to $E_\mathrm{fin}$. These particular parameters are used in inputs for our close coupling scattering calculations and their values were selected following systematic convergence test calculations with a maximum of 1\% mean deviation threshold criteria for $j_\mathrm{max}$ and 0.01\% for  $J_\mathrm{tot}$, respectively. Table~\ref{tab:conv-param-pH2} shows the corresponding parameters for $ortho$-H$_3$O$^+$ and $para$-H$_3$O$^+$ collision with $para$-H$_2$, while Table~\ref{tab:conv-param-oH2} provides those of for collision with $ortho$-H$_2$ projectile. Due to the higher ground rotational level of $ortho$-H$_2$, the maximum total energy considered for collision with this species is consequently higher.

\begin{table*}
	\centering
    \caption{The converged values of rotational basis size ($j_\mathrm{max}$) and maximum total angular momentum ($J_\mathrm{tot}$) parameters for the particular total energy intervals (from $E_\mathrm{init}$ to $E_\mathrm{fin}$) for the {\it o/p-}H$_3$O$^+$ -- {\it p-}H$_2$ collision. The step size for the energies ($E_\mathrm{step}$) in these intervals are also listed.}
    \label{tab:conv-param-pH2}
    \begin{tabular}{cccccc|ccccc}
        \hline
        \multicolumn{5}{c}{\textbf{\textit{ortho-}H$_3$O$^+ -$ \textit{para-}H$_2$ collision}} & &  \multicolumn{5}{c}{\textbf{\textit{para-}H$_3$O$^+ -$ \textit{para-}H$_2$ collision}} \\
        \hline
        $E_\mathrm{init}$  & $E_\mathrm{fin}$  & $E_\mathrm{step}$ & $j_\mathrm{max}$ & $J_\mathrm{tot}$ &  & $E_\mathrm{init}$  & $E_\mathrm{fin}$  & $E_\mathrm{step}$ & $j_\mathrm{max}$ & $J_\mathrm{tot}$ \\
        $[$cm$^{-1}]$ & $[$\cmmo$]$  & $[$\cmmo$]$ &  & & & $[$cm$^{-1}]$ & $[$\cmmo$]$  & $[$\cmmo$]$ &  &  \\
        \hline
        55.4 & 99.9 & 0.1 & 9 & 23 & &47.1 & 74.9 & 0.1 & 17 & 18 \\
        100 & 149.9 & 0.1 & 9 & 30 & &75 & 99.9 & 0.1 & 17 & 22 \\
        150 & 174.5 & 0.5 & 9 & 34 & &100 & 124.9 & 0.1 & 17 & 26 \\
        175 & 199.5 & 0.5 & 9 & 35 & &125 & 149.9 & 0.1 & 17 & 30 \\
        200 & 224.5 & 0.5 & 10 & 42 & &150 & 174.5 & 0.5 & 17 & 34 \\
        225 & 249.5 & 0.5 & 10 & 43 & &175 & 199.5 & 0.5 & 17 & 38 \\
        250 & 349.5 & 0.5 & 10 & 50 & &200 & 224 & 1 & 17 & 39 \\
        350 & 374.5 & 0.5 & 10 & 51 & &225 & 249 & 1 & 17 & 43 \\
        375 & 399.5 & 0.5 & 10 & 55 & &250 & 274 & 1 & 17 & 44 \\
        400 & 424.5 & 0.5 & 10 & 56 & &275 & 299 & 1 & 17 & 48 \\
        425 & 449.5 & 0.5 & 12 & 57 & &300 & 324 & 1 & 17 & 49 \\
        450 & 474.5 & 0.5 & 12 & 58 & &325 & 374 & 1 & 17 & 51 \\
        475 & 499.5 & 0.5 & 12 & 60 & &375 & 399 & 1 & 17 & 52 \\
        500 & 508 & 2 & 12 & 73 & &400 & 424 & 1 & 17 & 53 \\
        510 & 522.5 & 2.5 & 12 & 73 & &425 & 449 & 1 & 17 & 54 \\
        525 & 547.5 & 2.5 & 12 & 74 & &450 & 474 & 1 & 17 & 55 \\
        550 & 570 & 5 & 12 & 74 & &475 & 499 & 1 & 17 & 56 \\
        575 & 620 & 5 & 12 & 79 & &500 & 518.75 & 1.25 & 17 & 58 \\
        625 & 645 & 5 & 12 & 81 & &520 & 547.5 & 2.5 & 17 & 58 \\
        650 & 670 & 5 & 12 & 82 & &550 & 570 & 5 & 17 & 59 \\
        675 & 695 & 5 & 12 & 83 & &575 & 620 & 5 & 17 & 60 \\
        700 & 720 & 5 & 13 & 84 & &625 & 695 & 5 & 16 & 62 \\
        725 & 745 & 5 & 13 & 85 & &700 & 745 & 5 & 16 & 63 \\
        750 & 770 & 5 & 13 & 87 & &750 & 800 & 5 & 16 & 65 \\
		775	&	800	&	5	&	13	&	93	& &	810	&	820	&	10	&	16	&	66	\\
		810	&	900	&	10	&	13	&	94	& &	830	&	870	&	10	&	16	&	68	\\
		910	&	950	&	10	&	13	&	95	& &	880	&	920	&	10	&	15	&	70	\\
		960	&	980	&	20	&	13	&	95	& &	930	&	940	&	10	&	15	&	71	\\
		1000	&	-	&	50	&	13	&	98	& &	950	&	970	&	10	&	15	&	71	\\
		1050	&	-	&	50	&	13	&	106	& &	980	&	990	&	10	&	14	&	72	\\
		1100	&	1150	&	50	&	13	&	107	& &	1000	&	1020	&	20	&	14	&	73	\\
		1200	&	1250	&	50	&	13	&	109	& &	1040	&	-	&	20	&	14	&	75	\\
		1300	&	1350	&	50	&	13	&	112	& &	1060	&	1080	&	20	&	14	&	77	\\
		1400	&	-	&	50	&	13	&	120	& &	1100	&	-	&	50	&	14	&	78	\\
		1450	&	1500	&	50	&	13	&	121	& &	1150	&	-	&	50	&	13	&	78	\\
			&		&		&		&		& &	1200	&	-	&	50	&	13	&	80	\\
			&		&		&		&		& &	1250	&	-	&	50	&	13	&	84	\\
			&		&		&		&		& &	1300	&	-	&	50	&	13	&	89	\\
			&		&		&		&		& &	1350	&	-	&	50	&	13	&	91	\\
			&		&		&		&		& &	1400	&	-	&	50	&	13	&	93	\\
			&		&		&		&		& &	1400	&	1500	&	50	&	13	&	94	\\
        \hline
    \end{tabular}
\end{table*}

\begin{table*}
	\centering
    \caption{The converged values of rotational basis size ($j_\mathrm{max}$) and maximum total angular momentum ($J_\mathrm{tot}$) parameters for the particular total energy intervals (from $E_\mathrm{init}$ to $E_\mathrm{fin}$) for the {\it o/p-}H$_3$O$^+$ -- {\it o-}H$_2$ collision. The step size for the energies ($E_\mathrm{step}$) in these intervals are also listed.}
    \label{tab:conv-param-oH2}
    \begin{tabular}{cccccc|ccccc}
        \hline
        \multicolumn{5}{c}{\textbf{\textit{ortho-}H$_3$O$^+ -$ \textit{ortho-}H$_2$ collision}} & &  \multicolumn{5}{c}{\textbf{\textit{para-}H$_3$O$^+ -$ \textit{ortho-}H$_2$ collision}} \\
        \hline
        $E_\mathrm{init}$  & $E_\mathrm{fin}$  & $E_\mathrm{step}$ & $j_\mathrm{max}$ & $J_\mathrm{tot}$ &  & $E_\mathrm{init}$  & $E_\mathrm{fin}$  & $E_\mathrm{step}$ & $j_\mathrm{max}$ & $J_\mathrm{tot}$ \\
        $[$cm$^{-1}]$ & $[$\cmmo$]$  & $[$\cmmo$]$ &  & & & $[$cm$^{-1}]$ & $[$\cmmo$]$  & $[$\cmmo$]$ &  &  \\
        \hline
175.0	&	199.9	&	0.1	&	16	&	26	& &	165.9	&	174.9	&	0.1	&	16	&	22	\\
200.0	&	224.9	&	0.1	&	16	&	27	& &	175.0	&	199.9	&	0.1	&	16	&	29	\\
225.0	&	249.9	&	0.1	&	17	&	31	& &	200.0	&	249.9	&	0.1	&	16	&	36	\\
250.0	&	274.9	&	0.1	&	17	&	35	& &	250.0	&	274.9	&	0.1	&	16	&	38	\\
275.0	&	299.9	&	0.1	&	16	&	42	& &	275.0	&	299.9	&	0.1	&	16	&	45	\\
300.0	&	320.0	&	0.2	&	16	&	46	& &	300.0	&	319.8	&	0.2	&	16	&	52	\\
320.5	&	324.5	&	0.5	&	16	&	46	& &	320.5	&	349.5	&	0.5	&	16	&	56	\\
325.0	&	374.8	&	0.5	&	16	&	67	& &	350.0	&	500.0	&	0.5	&	16	&	60	\\
375.0	&	499.5	&	0.5	&	18	&	67	& &	502.5	&	547.5	&	2.5	&	16	&	60	\\
500.0	&	520.0	&	1	&	18	&	67	& &	550.0	&	597.5	&	2.5	&	17	&	64	\\
522.5	&	-		&	2.5	&	18	&	67	& &	600.0	&	647.5	&	2.5	&	17	&	68	\\
525.0	&	547.5	&	2.5	&	16	&	67	& &	650.0	&	697.5	&	2.5	&	17	&	75	\\
550.0	&	597.9	&	2.5	&	16	&	69	& &	705		&	820		&	5	&	16	&	75	\\
600.0	&	647.5	&	2.5	&	16	&	71	& &	825		&	870		&	5	&	16	&	77	\\
650.0	&	674.5	&	2.5	&	16	&	72	& &	875		&	920		&	5	&	16	&	80	\\
675.0	&	697.5	&	2.5	&	16	&	73	& &	925		&	945		&	5	&	14	&	80	\\
700		&	795		&	5	&	16	&	74	& &	950		&	995		&	5	&	14	&	87	\\
800		&	845		&	5	&	16	&	75	& &	1000		&	1090		&	10	&	14	&	89	\\
850		&	895		&	5	&	15	&	78	& &	1100		&	1120		&	10	&	14	&	90	\\
900		&	995		&	5	&	15	&	88	& &	1130		&	1140		&	10	&	13	&	91	\\
1000		&	1120		&	10	&	14	&	88	& &	1150		&	1200		&	10	&	13	&	96	\\
1130		&	1170		&	10	&	14	&	89	& &	1220		&	1260		&	20	&	13	&	96	\\
1180		&	1240		&	10	&	13	&	91	& &	1280		&	1320		&	20	&	13	&	103	\\
1250		&	1270		&	10	&	13	&	101	& &	1340		&	-		&	20	&	12	&	103	\\
1280		&	1290		&	10	&	12	&	103	& &	1360		&	1400		&	20	&	12	&	106	\\
1300		&	1350		&	50	&	12	&	103	& &	1450		&	1500		&	50	&	12	&	106	\\
1400		&	1400		&	50	&	11	&	103	& &	1550		&	1650		&	50	&	12	&	117	\\
1500		&	-		&	50	&	11	&	109	& &	1700		&	-		&	50	&	12	&	122	\\
1600		&	1700		&	50	&	11	&	110	& &			&			&		&		&		\\
        \hline
    \end{tabular}
\end{table*}


\bsp	
\label{lastpage}
\end{document}